\documentclass[a4paper]{jpconf}
\usepackage{graphicx}
\usepackage{iopams}
\begin{document}
\title{The QUAX proposal: a search of galactic axion with magnetic materials}

\author{G Ruoso$^1$, A Lombardi$^1$, A Ortolan$^1$, R Pengo$^1$, C Braggio$^2$, G Carugno$^2$, C S Gallo$^2$, C C Speake$^3$}
\address{$^1$ Laboratori Nazionali di Legnaro, Viale dell'Universit\`a 2, 35020 Legnaro (Italy) \\
$^2$ INFN and Dipartimento di Fisica e Astronomia, Via Marzolo 8, 35131 Padova  (Italy)\\
$^3$ School of Physics and Astronomy, University of Birmingham, West Midlands B15 2TT (UK)}
\ead{Giuseppe.Ruoso@lnl.infn.it}

\begin{abstract}
Aim of the QUAX (QUaerere AXion) proposal is to exploit the interaction of  cosmological axions with the spin of electrons in a magnetized sample. Their  effect is equivalent to the application of an oscillating rf field with frequency and amplitude which are fixed by axion mass and coupling constant, respectively.  The rf receiver module of the QUAX detector consists of magnetized samples with the Larmor resonance frequency tuned to the axion mass by a polarizing static magnetic field. The interaction of electrons with the 
axion-equivalent rf field produces oscillations in the total magnetization of the samples. To amplify such a tiny field,  a  pump field at the same frequency is applied in 
a direction orthogonal to the polarizing field.   The induced oscillatory magnetization along the polarizing field is measured by  a 
SQUID amplifier operated at its quantum noise level.

\end{abstract}

\section{Introduction}
An outstanding result of  modern cosmology is that a significant fraction of the universe is made of dark matter. However, the nature of such component is still unknown, apart its gravitational interaction with ordinary baryonic matter. A favored candidate for dark matter is the axion: a new particle introduced by Peccei and Quinn   to solve the strong CP problem \cite{PQ}, i.e. the absence of CP violation in the strong interaction sector of the Standard Model. Axions have properties similar to a $\pi_0$ particle and have mass $m_a$ inversely proportional to the Peccei-Quinn  symmetry breaking scale $f_a$.
For certain ranges of $f_a$ and $m_a$ (typically with masses ranging from $\mu$eV to meV), large quantities of axions may had been produced in the early Universe that could be able to account for a large portion of the cold dark matter forming today galactic halos. Axions have extremely small coupling to normal matter and radiation, but they can be converted into detectable photons by means of the inverse Primakoff effect as shown by Sikivie \cite{Sikivie}. The idea of Sikivie has been exploited by the several experiments \cite{MelissinosAX, Florida}, of which the most recent  is ADMX \cite{ADMX}. The latter experiment is still running, and for the moment it has been capable of exploring the axion model for masses of a few $\mu$eV. 

The QUAX (QUaerere AXion) proposal explores in details the ideas of Krauss \cite{Krauss1985}, Barbieri et al \cite{Barbieri1989}, and Kolokolov and Vorobyev \cite{Kolo1991}.  These authors proposed to study the interaction of the cosmological axion with the spin of fermions (electrons or nucleons). In fact, due to the motion of the Solar System through the galactic halo, the Earth is effectively moving through the cold dark matter cloud surrounding the Galaxy and an observer on Earth will see such axions as a wind. In particular, the effect of the axion wind on a magnetized material can be described as an effective oscillating rf  field with frequency determined by $m_a$, and amplitude related to $f_a$. Thus, a possible detector for the axion wind can be a magnetized sample with Larmor resonance frequency tuned to the axion mass by means of an external polarizing static field (e.g. 0.6 T for 17 GHz, corresponding to a 70 $\mu$eV axion mass). The interaction with the axion effective field will drive the total magnetization of the sample, and so producing oscillations in the magnetization that, in principle, can be detected. In order to optimize the detection, a pump rf field is applied in a direction orthogonal to the polarizing field. Due to the non linearity of Bloch equations in magnetized materials, a pump field applied to a suitable magnetized sample, (e.g. a small Yttrium Iron Garnet (YIG) sphere) can amplify the equivalent rf field generated by the axion wind. The induced change in the magnetic flux along the polarizing field is then fed to a SQUID magnetometer through a superconducting transformer. It is worth noticing that the magnetized sample must be cooled to ultra-cryogenic temperature to avoid fluctuations of the magnetization due to the thermal bath.  

The QUAX collaboration is exploring the experimental feasibility of the proposed detection scheme, starting at a precise value of the axion mass of about 70 $\mu$eV. 
The QUAX R\&D activities are conducted at the Laboratori Nazionali di Legnaro (LNL) of the Istituto Nazionale di Fisica Nucleare (INFN), 
and funded in the framework of the research call {\it What Next} of INFN.
We are building a prototype detector to study its sensitivity, intrinsic and technical noises, and demonstrate the possibility to reach the signal level corresponding to the axion coupling constant as predicted by the theoretical model.

\section{Theoretical introduction}

Let's start from the well known Lagrangian which describes the interaction of 1/2 spin particle with the axion field $a(x)$    
\begin{equation}\label{eq1}
L=\bar{\psi}(x)(i\hbar 
\gamma^\mu\partial_\mu- mc)\psi(x) - ia(x) \bar{\psi}(x)g_p\gamma_5\psi(x)
\end{equation}
where $\psi(x)$  and $a(x)$ are the spinor field of the fermion with mass $m$ and the axion field with dimension 
of momentum, respectively; here $\gamma^\mu$ are the 4 Dirac matrices, $\gamma^5=i \gamma^0\gamma^1\gamma^2\gamma^3$, and $a(x)$  is coupled to the matter by the  pseudo-scalar coupling constant $g_p$. By obtaining the Euler-Lagrange equation and in the non relativistic limit,  the time evolution of a 1/2 spin particles can be described by the usual Schroedinger equation
\begin{equation}\label{eqsh}
i\hbar\frac{\partial \varphi}{c\partial t}=\left[-\frac{\hbar^2}{2m}\nabla^2+g_sca- \frac{g_p\hbar }{2m}\boldsymbol{\sigma}\cdot \boldsymbol{\nabla}a \right]\varphi\ , 
\end{equation}
where the most relevant interaction term 
\begin{equation}
 -\frac{g_p\hbar }{2m}\boldsymbol{\sigma}\cdot \boldsymbol{\nabla}a \equiv -2 \frac{e \hbar }{2m}\boldsymbol{\sigma}\cdot\left(\frac{g_p}{2e}  \right)\boldsymbol{\nabla}a
\end{equation} 
has the form of the interaction between the spin magnetic moment of a fermion   ($-2 \frac{e \hbar }{2m}\boldsymbol{\sigma} = -\mu_B \boldsymbol{\sigma}$ with $\mu_B$ the Bohr magneton) and an effective 
 magnetic field $B_a \equiv \frac{g_p}{2e}  \boldsymbol{\nabla}a $ .  It is worth noticing that the axion field couples to fermions as a pseudo-scalar field; in fact, $\boldsymbol{\nabla}a$  is a pseudo-vector (like the magnetic field) and $\boldsymbol{\sigma}\cdot \boldsymbol{\nabla} a$ is a true scalar interaction term. 

Axions represent the best example of non-thermal dark matter candidate \cite{Turner}. 
The expected dark matter density is $\rho\simeq 300$ MeV/cm$^3$, and we will suppose that axion is the dominant component. As the axion mass should be in the range
 $ 10^{-6}{\rm eV} < m_a < 10^{-2} {\rm eV}$, we have $n_a\sim 3\times10^{12}\  (10^{-4}\ {\rm eV}/m_a)$ axions per cubic centimeter which is a remarkably high density number. The axion velocities $v$ are distributed according to a Maxwellian distribution (the non relativistic limit of the Bose-Einstein distribution),
 with a velocity dispersion
$\sigma_v\approx 270$ km/sec.
Due to Galaxy rotation and Earth motions in the Solar system, the rest frame of an Earth based laboratory is moving through the local axion cloud with a time varying velocity $\boldsymbol{v}_E= \boldsymbol{ v}_S+ \boldsymbol{ v}_O+ \boldsymbol{ v}_R$, where  $\boldsymbol{ v}_S$  represents the Sun velocity in the galactic rest frame (magnitude 230 km/sec), $\boldsymbol{ v}_O$  is the Earth's orbital velocity around the Sun (magnitude 29.8 km/sec), and $ \boldsymbol{ v}_R$ the Earth's rotational velocity (magnitude 0.46 km/sec).  
The observed axion velocity is then $\boldsymbol{ v}_a = \boldsymbol{ v}- \boldsymbol{ v}_E$ . The effect of this motion is to broaden the Maxwell distribution, as well as to modulate it with a periodicity of one sidereal day and one sidereal year. 

Axions kinetic energy  is expected to be distributed with a mean relative to the rest mass  $7 \times 10^{-7}$ and a dispersion about the mean  $5.2 \times 10^{-7}$ \cite{turner1}. The inverse of this last number represent the natural figure of merit of the axion linewidth, $Q_a \simeq 1.9\times 10^6$.
The mean De Broglie wavelength of an axion  is
$\lambda_d\simeq {h}/ ({m_a v_a}) \simeq  13.8 \left({10^{-4} {\rm eV}}/{m_a} \right)$ m,
 therefore $\lambda_d$ is much greater than the typical length of experimental apparatus, in our case the magnetized samples. 
 Such theoretical and experimental aspects allow us to  treat $a(x)$ as a classical field that interacts coherently with  fermions with a mean value 
$a(x)=a_0 {\rm exp}[ {i ( p^0 c t-\boldsymbol{p}_E\cdot\boldsymbol{x} )/\hbar}]$
where $\boldsymbol{ p}_E = m_a \boldsymbol{ v}_E$, $p^0=\sqrt{m_a^2 c^4 + |\boldsymbol{ p}_E |^2 c^2} \approx m_a c^2 + |\boldsymbol{p}_E|^2 /(2 m_a)$ and $a_0$ is the field amplitude.
The amplitude $a_0$ can be easily computed by equating the moment transported by this field per unit of volume (i.e. the associated energy momentum tensor $T^{0i}=a_0^2 p^0 p_E^i$) to the number of axion per unit volume times the average momentum (i.e. $n_a<p^i> =n_a p_E^i$), and it reads 
$a_0=\sqrt{({n_a\hbar^3 c})/({m_a c}})$.
To calculate the effective magnetic field associated with the mean axion field we multiply the gradient $\boldsymbol{\nabla}a(x) =i (\boldsymbol{p}_E / \hbar)\  a(x)$ by $g_p/(2 e)$ and take the real part
 \begin{equation}
\mathbf{B}_a = \frac{g_p}{2 m \gamma}  \left(\frac{n_a\hbar }{m_a c}\right)^{1/2}\, \boldsymbol{ p}_E \,  \sin\left(\frac{p^0 c t-\boldsymbol{p}_E\cdot\boldsymbol{x}}{\hbar}\right)
\end{equation} 
where $\gamma\equiv g_L e/ (2 m)$ is the gyromagnetic ratio and  $g_L=2$ is the Land\'e g-factor for elementary fermions.  
In the framework of DFSZ model of axions\cite{DFSZ,DFSZ1}, the value of the coupling constant $g_p$ with electrons  can be calculated and expressed in terms of  the $\pi^0$ mass and decay constant
 $m_{\pi^0}=135 \, {\rm MeV}$ and  $f_{\pi^0} = 93 \, {\rm MeV}$ as  
$g_p \simeq ({m \, m_a})/({m_{\pi^0} f_{\pi^0}}) = 2.8 \times 10^{-11} ({m_a}/({1\  \rm{eV}}))  $
where we have used $ m =0.5  \,{\rm MeV}$ for the electron mass. 

Putting the magnetized samples in $\boldsymbol{ x} =\boldsymbol{ 0}$, we have that the equivalent oscillating rf field along the $\boldsymbol{ p}_E$ direction has a mean amplitude and central frequency 
\begin{equation}
B_a  = 9.2 \cdot 10^{-23} \left(\frac{m_a}{10^{-4} {\rm eV}}\right)    \,\,\,\, {\rm T}\ ,\,\,\,\,\, \frac{\omega_a}{2 \pi}  =  24  \left(\frac{m_a}{10^{-4} {\rm eV}}\right)    \,\,\,\, {\rm GHz} \, 
\label{axionfield}
\end{equation}
with a relative linewidth $\Delta \omega_a/\omega_a \simeq 5.2 \times 10^{-7}$.  

\section{Experimental scheme}
To detect an extremely small  rf field we will exploit the Electron Spin Resonance (ESR)  in a magnetized media.
 In particular, we have to recourse to the continuous wave ESR in  a magnetic material using a coil oriented along the polarizing field direction.  
This method is known as  LOngitudinal Detection scheme (LOD), and during the last decades it was investigated systematically, 
both theoretically and experimentally \cite{pescia}. 

A magnetized spherical sample of volume $V_s$ and magnetization $M_0$  is placed  in the bore of a solenoid, 
which generates a static magnetic field $B_0$ (polarizing field). 
The value $B_0$  determines the Larmor frequency of the electrons, 
and so the axion mass  under scrutiny, through the relation  
\begin{equation}
B_0 = \frac{\omega_L} {\gamma} = \frac{ m_a c^2}{\gamma \hbar } = 0.85 \left(\frac{m_a}{10^{-4} {\rm eV}}\right)  \,\,\,\, {\rm T}.
\end{equation}
Then, an additional radio frequency field  $B_{1}$  is applied to the sample in a direction orthogonal to $\mathbf{B}_0$, that we suppose along the $z$ axis.
As the radiofrequency field is in the $x-y$ plane, it can drive 
 the magnetic resonance causing the electron spin to flip between the two Zeeman sublevels. We don't need the quantum formalism as the rf field amplitude 
is sufficiently large to drive a macroscopic number of spins. The evolution of 
magnetization $\mathbf{ M}$  of  the sample is described by the  Bloch equations with dissipations  and radiation damping \cite{bloom}
\begin{eqnarray}\label{bloch}
\frac{dM_x}{dt}&=&\gamma  (\mathbf{M}\times\mathbf{ B})_x - \frac{M_x}{\tau_2} -\frac{M_x M_z}{M_0 \tau_r} \nonumber \\
\frac{dM_y}{dt}&=&\gamma (\mathbf{M}\times\mathbf{ B})_y - \frac{M_y}{\tau_2} - \frac{M_y M_z}{M_0\tau_r} \nonumber \\
\frac{dM_z}{dt}&=&\gamma (\mathbf{M}\times\mathbf{ B})_z - \frac{M_0- M_z}{\tau_1}- \frac{M_x^2+M_y^2}{M_0 \tau_r}
\end{eqnarray}

where $M_0$ is the static magnetization directed along the $z$ axis, $\tau_r$ is the radiation damping time, $\tau_1$ and $\tau_2$ are the longitudinal (or spin-lattice) and transverse (or spin-spin) relaxation time, respectively.  The non linear terms proportional to $\tau_r^{-1}$ were introduced by Bloom in 1957 to account for electromotive force  induced in the rf coils of the driving circuit by magnetization changes without 
taking into account the dynamics of the rf coils. In this case  $ \tau_r^{-1}= 2 \pi \zeta \gamma \mu_0 M_0 Q$, where the filling factor $\zeta$ and the quality factor $Q$  account for  geometrical coupling of coil and magnetized sample and  circuit dissipations, respectively.  

 In the  higher frequency regime, which is our case,  $\omega_L>  \widehat{\omega}_L $, radiation damping is dominated by magnetic dipole  emission in free space from the magnetized sample,  and  
the relaxation time  reduces  to  $ \tau_r^{-1}=1/(4 \pi) \frac{\omega_L^3}{c^3}  \gamma \mu_0 M_0 V_s $. 
The steady state solutions of  Eq.(\ref{bloch}) in the presence of radiation damping and for various approximations  are given in ref. \cite{augustine}.  
A new relaxation time is introduced $\tau^\star_2=1/(\tau_2^{-1} + \tau_r^{-1})$.

Let us suppose that the rf field is linearly polarized and a linear superposition of the equivalent field  $\mathbf{ B}_a$, due to the axions, and a pump field $\mathbf{ B}_p$, 
with amplitude $B_p>>B_a$. We assume also that the axion field direction is parallel to the direction of $\mathbf{B}_a$. In addition, the axion and pump frequencies $\omega_a$ and $\omega_p$ are within the linewidth of the Larmor resonance  of the magnetized sample, i.e. ($\omega_p-\omega_L)<1/\tau_2^\star$ and 
 ($\omega_a-\omega_L)<1/\tau_2^\star$. In this way, the Larmor frequency of electrons is driven by  the rf field 
\begin{equation}
\label{fields}
B_1 = B_p \cos \omega_p t + B_a \cos \omega_a t .
\end{equation}

Due to the non-linearity of the Bloch's equations, as a result of this {\it amplitude modulated} rf field, a variable magnetization is produced along the $z$-axis.  
In fact,  the stationary solution for $M_z$ in Eq.(\ref{bloch}) can be used  even in quasi-stationary regimes \cite{augustine}, provided that $|\omega_p-\omega_a|<<\omega_L$.  
At the beat note between the two field components  we have 
\begin{equation}
\Delta m_z\equiv M_z-M_0  \simeq \frac{M_0}{4}\frac{\tau_2^\star}{ \tau_2} \ \gamma^2 \tau_1\tau_2^\star  B_p B_a \cos \omega_D t\\
 \end{equation}
where we have defined the detection frequency $\omega_D=\omega_p-\omega_a$ and assumed $\omega_D<\min\{1/\tau_1,1/\tau_2^\star \} $. 

Such  low frequency oscillations of the magnetization can be detected with suitable devices, e.g. a SQUID amplifier. 
In this detection scheme we can choose the best detection frequency for increasing the signal to noise ratio, and we can use amplifiers that could not work 
at very high frequencies. Moreover, the effective rf field $B_a$ is down converted to a low frequency field with amplitude $B_D=\mu_0 \Delta m_z \equiv G_m B_a$, where $G_m$ is the transduction gain,  and it is given by
\begin{equation}
G_m= \mu_0 \frac{M_0}{4}\frac{\tau_2^\star}{\tau_2} \ \gamma^2 \tau_1\tau_2^\star  B_p
\end{equation}
If we assume that the relaxation times satisfy  $\tau_1 \sim \tau_2$   and $\tau_r<\tau_2$,  
 the  transduction gain will depend only on $\tau_r$:  $G_m \simeq \mu_0 M_0 \tau_r^2  \gamma^2  B_p \approx 1/(8 \pi^2) (\lambda_L^3/V_s) \gamma \tau_r  B_p  $, where $\lambda_L\equiv 2 \pi c /\omega_L $ is the wavelength corresponding to the Larmor frequency. 
 Thus to obtain a gain  $G_m>1$ in free space,  the sample volume must satisfy the inequality $V_s<  1/(8 \pi^2)\lambda_L^3 \tau_r \gamma B_p$. On the other hand, the pump field amplitude must be far
 from saturation  $\gamma^2 B_p^2 \tau_r \tau_1 <<1$, which implies $\gamma \tau_r  B_p <<1$.  This is the reason why we can't get $G_m>1$ in free space with realistic sample volumes and pumping fields.  

Inside an rf cavity, this problem could be solved  
but the thermal noise in the hybridized system formed by cavity and magnetized sample is much greater than the 
axion equivalent field. 
The solution is to detect axions by placing the sample in a 
waveguide  with a cutoff frequency $\omega_c$ above the Larmor frequency of the sample. 
For instance, we can place  the sample  at a distance $\ell$ from the aperture of  a rectangular waveguide of cross section $ab$.
The lowest cut-off frequency for an open waveguide $(a>b)$ is given by (the waveguide behaves as a high pass filter):
$\omega_c={1}/({4 \pi a \sqrt{\epsilon \mu}})$.

Due to boundary conditions,  if the Larmor frequency is lower than $\omega_c$,  the magnetic resonant mode cannot propagate inside 
the waveguide, but only evanescent waves can exist. We expect then a reduction of the radiation damping mechanism for the 
selected magnetic resonance.  
The advantage of this experimental configuration is twofold: i) relaxation times are no 
longer dominated by radiation damping but by the intrinsic relaxation times of the sample 
$\tau_1$ and $\tau_2$; and ii) the noise associated to thermal photons is greatly reduced. 
The main difference with the case of a microwave cavity is that we exploit  the non-radiative field to pump 
energy in the sample by means of a rf coil placed around the sample itself.

\section{Noise considerations}

We are interested in  fluctuations  of the magnetization of the sample. There are at least four sources of noise to consider:
a) pump field noise;
b) noise at the Axion frequency due to thermal photons that can be down converted after interaction with the pump field;
c) magnetization noise generated by sample that can either be down converted by the pump or can act as a noise source in the low frequency readout;
d) noise associated with the SQUID pickup at the readout frequency.
A preliminary evaluation of these noise sources shows that the most challenging issue is related with the pump field noise. All the noises are in the process of being carefully evaluated.

In addition to these noise sources, we have to take into account any relaxation process that can be active at the readout frequency and 
can induce fluctuations in the z component of magnetization. The presence of such processes and their magnitude has 
to be directly measured on the magnetized sample.

\end{document}